\documentclass[10pt,twocolumn,letterpaper]{article}
\usepackage{pdfpages}
\usepackage{url} % for url
\usepackage[bookmarks=false]{hyperref} % for citations and figure references
\usepackage{graphicx}
\usepackage{amsmath}
\usepackage{amssymb} % for mathbb
\usepackage[utf8]{inputenc}
\usepackage[english]{babel}
\usepackage{csquotes}
\usepackage{xr}
\usepackage{verbatim} % for comments
\usepackage{float} % for specifying *here* - "\begin{figure}[H]" images
\usepackage{cite}
\usepackage{fancyhdr}
\usepackage{geometry}
\usepackage{afterpage}

\usepackage{iccv}
\usepackage{times}
\usepackage{epsfig}

\iccvfinalcopy %

\def\iccvPaperID{0} %

\newcommand{\omt}[1]{}

\begin{document}

\title{ Designing the Next Generation of Intelligent Personal Robotic Assistants for the Physically Impaired }

\author{
	\small
	\begin{tabular}{c c c c}                              
		\bf Basit Ayantunde &
		\bf Jane Odum &
		\bf Fadlullah Olawumi &
		\bf Joshua Olalekan \\                            
		\multicolumn{4}{c}{University of Ilorin} \\                                              
		\multicolumn{4}{c}{\{rlamarrr, janeodum41, fqrious+god, olajoshua.tunji\}@gmail.com} \\
	\end{tabular}                                                                       
}

\date{}
\maketitle

\providecommand{\ckeywords}[1]{\textbf{\textit{Keywords:}} #1}

\ificcvfinal\thispagestyle{plain}\fi

\begin{abstract}
The physically impaired commonly have difficulties performing simple routine tasks without relying on other individuals who are not always readily available and thus make them strive for independence. While their impaired abilities can in many cases be augmented (to certain degrees) with the use of assistive technologies, there has been little attention to their applications in embodied AI with assistive technologies. This paper presents the \emph{modular framework, architecture, and design} of the \emph{mid-fidelity prototype} of \emph{MARVIN}: an artificial-intelligence-powered robotic assistant designed to help the physically impaired in performing simple day-to-day tasks. The prototype features a trivial locomotion unit and also utilizes various state-of-the-art neural network architectures for specific modular components of the system. These components perform specialized functions, such as automatic speech recognition, object detection, natural language understanding, speech synthesis, etc. We also discuss the constraints, challenges encountered, potential future applications and improvements towards succeeding prototypes.\\
\end{abstract}

\begin{ckeywords}  embodied AI, automatic speech recognition, object detection, natural language understanding, speech synthesis.
\end{ckeywords}

\section{Introduction}
There have been tremendous discoveries and breakthroughs in the field of artificial intelligence and robotics in the past few decades, and technology has accordingly grown from being seen as eliminating the need for humans to augmenting our abilities and in the process, making life easier. Consequently, this poses a challenge in bringing these technologies to their useful applications in our daily lives.

A reported estimate of 1 billion people live with one or more disabilities around the globe, in accordance with a population estimate of 9.6 billion people \cite{world2011world}. The aged also possess slowly degrading physical functionalities and functional impairments which vary from one individual to another \cite{VitaminD}. The global population aged 60 years and above were estimated to be 962 million in 2017 and expected to double by 2050, during which it is projected to reach nearly 2.1 billion \cite{unnAgeReport}.

These impairments make performing day-to-day tasks difficult for these individuals and thus affect the quality of their lives, i.e. hygiene, feeding, alimentations, work, etc \cite{VALENCA2017}. The aged phase of life is often made more difficult due to social factors such as subjectivity to discrimination, constructed social exclusion and stigmatization \cite{AgeStigma}.

Physical impairments can be caused by natural aging, diseases (hypertension, glaucoma, cataracts, etc.), accidents, injuries, natural impairments since birth, etc.

While assistive technologies addressing some of these physical impairments have been undergoing rigorous research and development for decades, they mostly take the form of aids and are often incorporated or attached to the impaired individual's body, which in many cases does not suffice well to enable them to perform their daily activities effectively. Some previous assistive technologies include electric wheelchairs, electroencephalography, digital walking sticks, bone-anchored hearing aid \cite{wiki:baha}, mobility scooters \cite{wiki:mobscooter}, cochlear implants \cite{wiki:cochlear_implant}, prosthesis \cite{wiki:prosthesis}, text-to-voice wands, intelligent personal digital assistants \cite{wiki:ipda}, etc.

Current IPDAs such as \emph{Alexa}, \emph{Siri}, and the \emph{Google Assistant} are not readily integrable and customizable to run on specialized hardware especially due to the closed-source natures of their architectures and software systems. Other challenges include immobility, non-interactivity with the physical world around them, and the few user interfaces they readily support leading to inaccessibility for people with the counter-impairments (relative to the user interface).

Our approach tries to help them perform some of these tasks. We propose a simple and modular robotic system utilized toward this end. We also propose an Intelligent Personal Robotic Assistant (IPRA) architecture, which in contrast to the Intelligent Personal Digital Assistant architectures (IPDA) \cite{ipdaArch}, is highly modular, readily integrable with various hardware components for robotics applications, and highly prioritizes the modular, multimodal and fusible nature of the user interfaces.

While the current prototype of the robotic system is basic, does not yet provide support for dynamic environments and specialized use cases, this only serves as a mid-fidelity prototype and at the same time, a proof-of-concept for our framework and architecture, and serves as a basis for future prototypes. The system can be easily built on, improved and used in coordination with pre-existing assistive technologies.

Our current system prototype utilizes a mixture of lightweight state-of-the-art neural network architectures towards a problem decomposition approach, which in turn makes it more reliable in comparison to fully End-to-End (E2E) AI robotic systems that are in many cases questionable due to the non-interpretability of the representations of the world around them, thus making their decisions highly difficult to probe.

Running modern state-of-the-art neural network architectures often requires huge computation power and large memory requirements due to the hundreds of millions and often billions of parameters they possess and in effect need to be deployed on costly hardware devices to meet real-time requirements.

While it is noteworthy that for the affordability and general acceptability of these systems, there is an impending need for them to be deployable on cost-effective hardware platforms often consequently possessing constrained resources as these cost money. This is addressed mostly by building a series of prototypes, and balancing resource consumption against the quality of solutions provided by these systems \cite{DBLP:journals/corr/abs-1906-08172} whilst paying attention to cost. Nonetheless, a common practice is to start with a simple system with optimal resources, incrementally add more features and accordingly add more resources as required. 

Also worth noting is the nature of complexity that arises from designing these systems, due to the non-deterministic timing of events often received by the system and real-time constraints frequently imposed in the process. In a lossless data processing pipeline, backpressure is always avoided and monitored, this is worsened when the designated hardware doesn't have enough resources to counteract the effects of these liable limitations. Our framework provides a few basic components to address these complexities and limitations.

The proposed architecture also highly prioritizes the modular, multimodal, and fusible nature of the user interfaces to enable it to serve a wide variety of people with/without disabilities, which most current IPDA architectures do not prioritize but instead possess non-modular unimodal user interfaces with limited interactivity and accessibility. The framework's flexibility also allows for the easy deployment of engineered, E2E, and reinforcement learning models within the architecture. The multimodal nature and fusible user interfaces allow the system to fuse and contextualize semanticized sampled data such as visual cues, gestures, and speech into higher-level semantic data. Another benefit of its multimodal nature is that it gives the system a better understanding of context.

\section{Capabilities}
\label{sec:capab}
Listed below are the assistive capabilities (\texttt{Skills}) implemented into the current prototype of the robot:
\begin{itemize}
	\item Facial Recognition
	\item Time Querying
	\item Schedule Organization
	\item Phone Call Initiation
	\item Object Finding
	\item People Finding
	\item Intruder Detection
	\item Visually Impaired Guidance
	\item Smart Home Devices Control (Bulbs, Fans, etc.)
\end{itemize}

\begin{figure*}[ht]
	\centering
	\setlength
	\fboxsep{0pt}
	\setlength
	\fboxrule{0.25pt}
	\includegraphics[width=0.75\textheight]{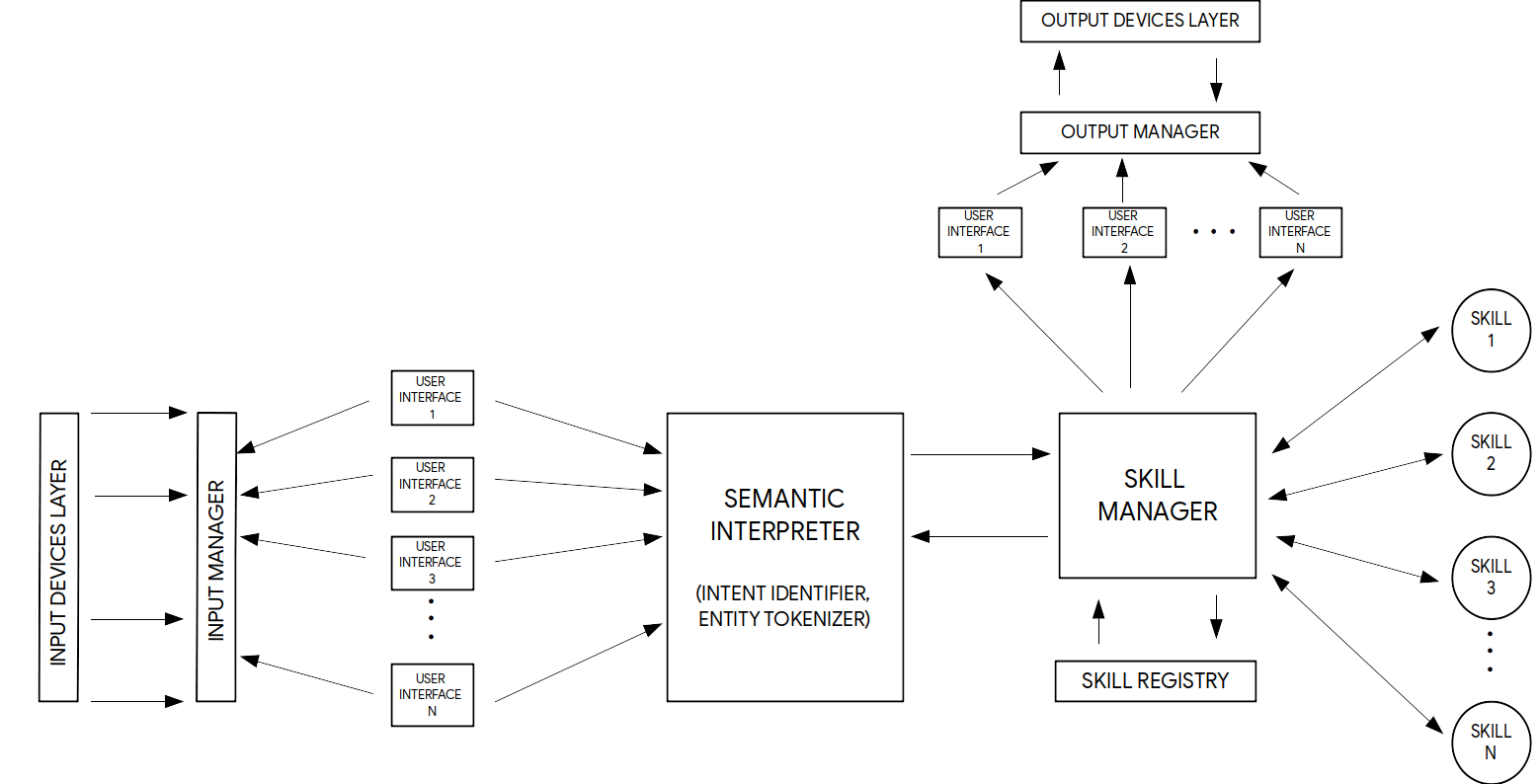}
	\caption{IPRA Architecture}
	\label{Arch}
\end{figure*}

\section{Software Architecture} \label{section:swa}
This section discusses the IPRA architecture's layers. The architecture diagram is shown in Figure (\ref{Arch}).

\subsection{Input/Output Devices Layer}
The system often needs to get/send data samples from/to the environment. The raw data samples are obtained from hardware components that are accessed/interfaced via the \texttt{Input/Output Devices Layer}. This is readily available within the Linux kernel (implemented via \textit{udev}) with various drivers that manage the hardware and resources used in the system.

\subsection{Input/Output Manager}
Various data samples can be received or sent via the \texttt{Input/Output Devices Layer} and they are dispatched by the \texttt{Input/Output Manager} to the appropriate \texttt{User Interfaces}. This is beneficial especially when there are multiple \texttt{User Interfaces} in use and a need to share or route the incoming raw data samples to the appropriate \texttt{User Interfaces} arises.

\subsection{User Interface}
The data samples collected from the environment or generated by the system often need various additional semantics for its utilization. For example in a voice interactivity setting, it is often required to merge various incoming audio samples from microphone arrays into a frame containing multi-channel samples. This is also useful in a setting where we need to merge/fuse camera and microphone samples (as in \cite{jarvis}). The \texttt{User Interface} performs this task and can take the form of a neural network, perform simple signal processing, and chained together to obtain different levels of semantics.

\subsection{Semantic Interpreter} \label{sub:sem}
The semanticized data samples are interpreted by the \texttt{Semantic Interpreter} into commands represented as \texttt{Skill} identifiers and its \texttt{Entities}. The \texttt{Skill} identifiers and its \texttt{Entities} are then passed to the \texttt{Skill Manager}. The \texttt{Semantic Interpreter} can be implemented as an Automatic Speech Recognizer, Gesture Recognizer, Sign Language Recognizer, etc. The implementation choice depends on the nature of the incoming data and end-user interface.

\subsection{Skill Manager}
The recognized \texttt{Skills} and \texttt{Skill} \texttt{Entities} are dispatched and evaluated by the \texttt{Skill Manager}. The \texttt{Skill Manager} also ensures the \texttt{Skill}'s \texttt{Entities} and requirements are completed and fulfilled by interacting with the user via an output \texttt{User Interface} composition (if any).

\subsection{Skill Registry} \label{skillRegArch}
The \texttt{Skill Registry} contains the \texttt{Skills'} definitions and requirements.

\subsection{Skill} \label{skillArch}
The \texttt{Skill} unit defines how the robot is meant to execute the semanticized instruction received through the \texttt{User Interface}. This can be developer or end-user defined via a generic though restricted interface. \texttt{Skills} are similar to Alexa Skills and Google Assistant Actions.

\section{Framework} \label{section:framework}
In this section, we describe the framework and its components used towards complementing the proposed IPRA architecture.
Our framework increases the speed of developing these systems and also makes them easier to conceptualize since they perform lots of real-time processing and involve complex data-flows often in a multi-threaded fashion. The framework implements the IPRA architecture introduced in Section \ref{Arch}.

These components are loosely coupled to allow for flexibility, composability, and suitability for unit testing.
The system is represented as a directed \textbf{graph} consisting of \texttt{Nodes} with multi-input/output channels termed \emph{Streams}. Some of the implemented components are based on some real-time design patterns such as \emph{Observer} and \emph{Watchdog}.

\subsection{Node}
A \texttt{Node} is the unit of the graph that performs computations. Not to be confused with ROS \texttt{Nodes} \cite{ros:nodes}, this can contain a whole program and does not rely on inter-process communication. This is highly similar to Mediapipe's \texttt{Calculators} \cite{DBLP:journals/corr/abs-1906-08172}. A \texttt{Node} runs on at most one thread at an execution timepoint. \texttt{Nodes} operating on different threads can communicate via \texttt{Streams}.

\subsection{Packets}
The \texttt{Packet} is a basic data unit of a generic type that flows from one \texttt{Node} to another. It is implemented as a cheaply and efficiently copied data structure.

\subsection{Streams}
Components of the framework are connected (send data) via \texttt{Streams}. \texttt{Streams} serve as a multi or mono output channel, therefore either serving one or more \texttt{Packets} per execution timepoint.
A \texttt{Stream} of \texttt{Packets} is implemented as a single-ended thread-safe queue.
\texttt{Streams} can be lossy or lossless. Lossy \texttt{Streams} can miss a specified number of successive \texttt{Packets} in case it is unable to keep up with the data flow rate and thus avoiding backpressure. Lossless \texttt{Streams} are not allowed to miss any \texttt{Packets} and must be processed within a specified time frame.

\subsection{Watchdog}
The \texttt{Watchdog} monitors the rate of data flow from one component to another along a \texttt{Stream}, ensures real-time constraints are met and monitors backpressure along the \texttt{Stream}. This is configured via a data-only structure specifying the requirements i.e latency, and throughput.

\subsection{Latch}
This is a mechanism that helps control the flow of \texttt{Packets} along the \texttt{Stream} and is useful when serving multiple \texttt{Streams} of which there is an impending need to control the flow of \texttt{Packets} to some or all of the \texttt{Streams}. The \texttt{Latch} is controlled via a binary-state input \texttt{Stream} that stops or resumes an already stopped flow of \texttt{Packets} along a \texttt{Stream}.

\subsection{Aggregator}
This component of the framework continuously slices accumulated sampled data across a specified interval (time-window), this is especially useful in aggregating audio or video samples for performing deep learning inference.

\subsection{Attention Node}
\texttt{Attention Nodes} are a special type of \texttt{Node} that serves as a state \texttt{Stream} to the system, this is useful when it is not necessary to perform actions on every received input. An \texttt{Attention Node} is implemented as a \texttt{Node} whose input parameters are generic input \emph{Streams} and outputs are \emph{Streams} of bits. This can be integrated with Deep Neural Networks and other algorithms.

\subsection{Skill Registry}
This framework component implements the \texttt{Skill Registry} component of the IPRA architecture (See Section \ref{skillRegArch}). A \texttt{Skill Registry} serves as a router from one data type to another, this includes integers, strings, pointers, function pointers, functors, etc.
For example, a \texttt{Skill Registry} is implemented as a hash table of string keys to functor values which are executed on the same thread or another thread depending on the specified execution policy.

\subsection{Skill}
This framework component complements the \texttt{Skill} component of the IPRA architecture (See Section \ref{skillArch}). A \texttt{Skill} can be as simple as a query-only \texttt{Skill} which only needs to respond to a user via the selected interface (e.g. speech synthesizer) upon request. Skills are of two types, namely:

\begin{itemize}
	\item High-Level Skills: Both developers and end-users can create this type of \texttt{Skill}. They only have access to the abstracted utilities and device platform's features exposed to them by the developer.
	
	\item Low-Level Skills: This type of \texttt{Skill} is intended for developer utilization. They typically have access to the lower level system utilities and hardware resources.\\
	
\end{itemize}

\begin{figure*}[ht]
	\centering
	\setlength\fboxsep{0pt}
	\setlength\fboxrule{0.25pt}
	\includegraphics[width=0.34\textheight]{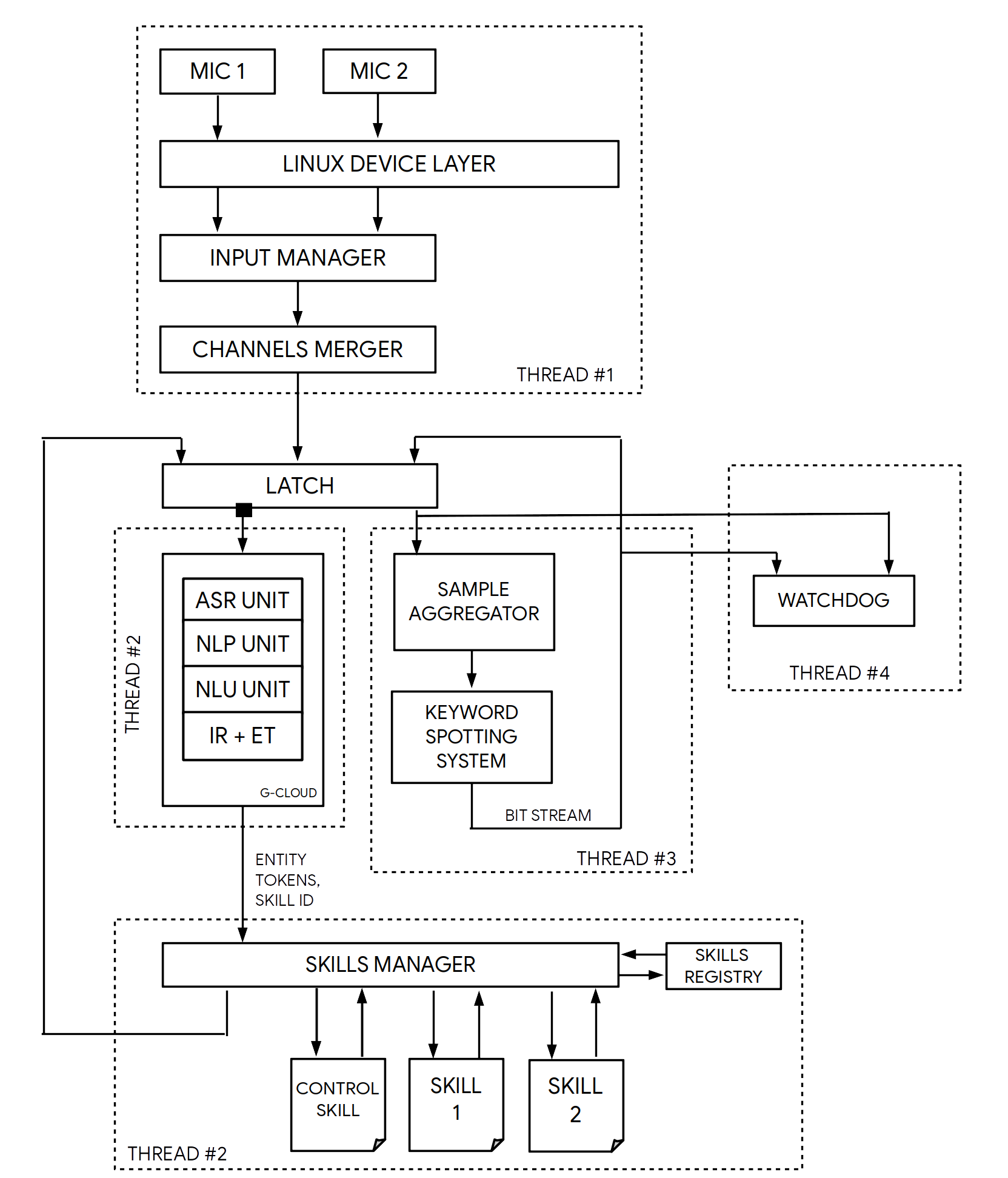}
	\caption{Speech Only Interface Architecture using our IPRA framework. This is a streamlined version of our original deployed architecture. The diagram highlights the key components of our framework used in implementing the architecture.}
	\label{KwsArch}
\end{figure*}

\section{Mechanical Design} \label{section:mech}
In this section, we describe the key mechanical units of the first Marvin prototype that enables it to interact with the physical world.

\begin{figure*}[ht]
	\centering
	\setlength\fboxsep{0pt}
	\setlength\fboxrule{0.25pt}
	\includegraphics[height=0.3\textheight]{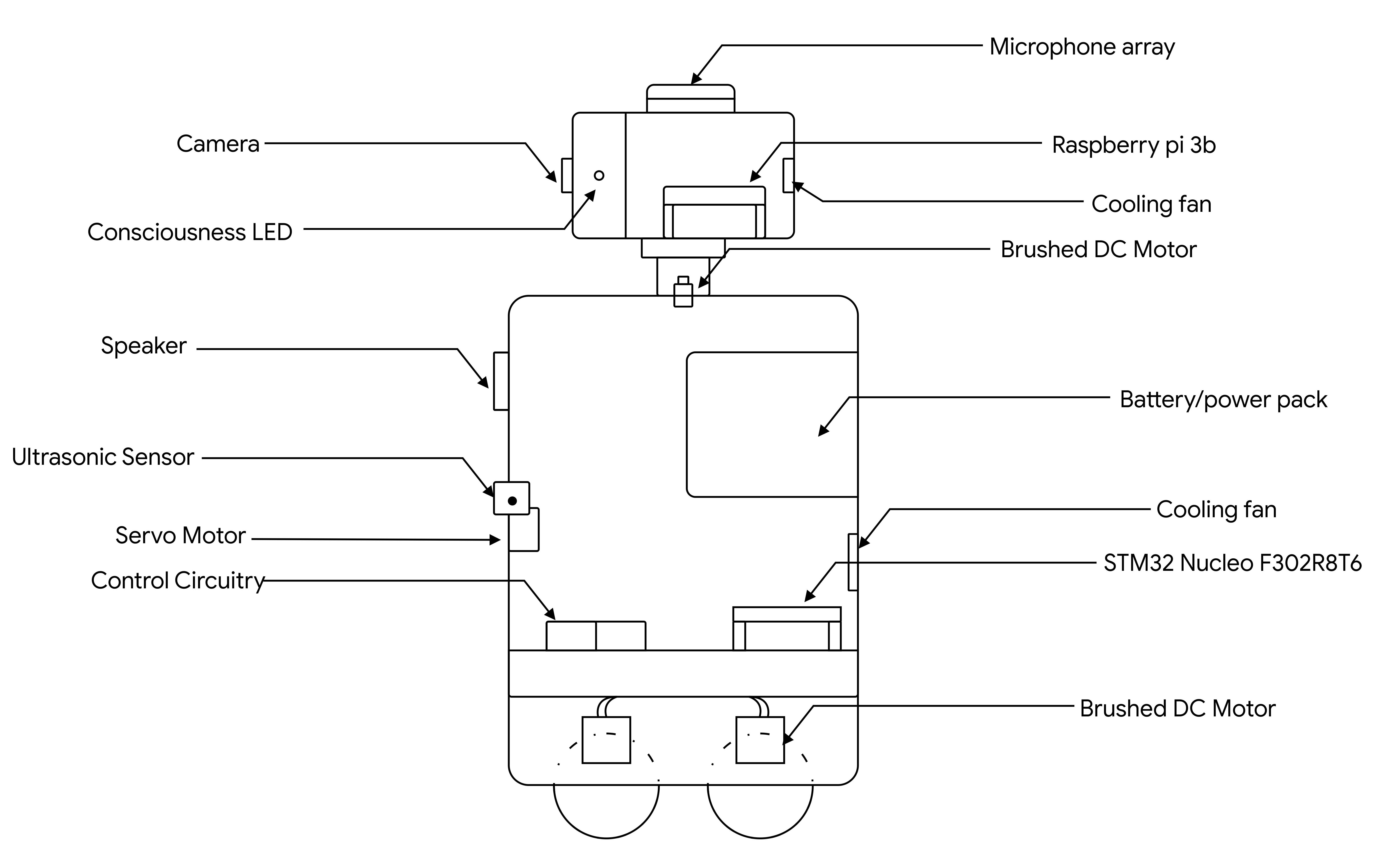}
	\includegraphics[height=0.3\textheight]{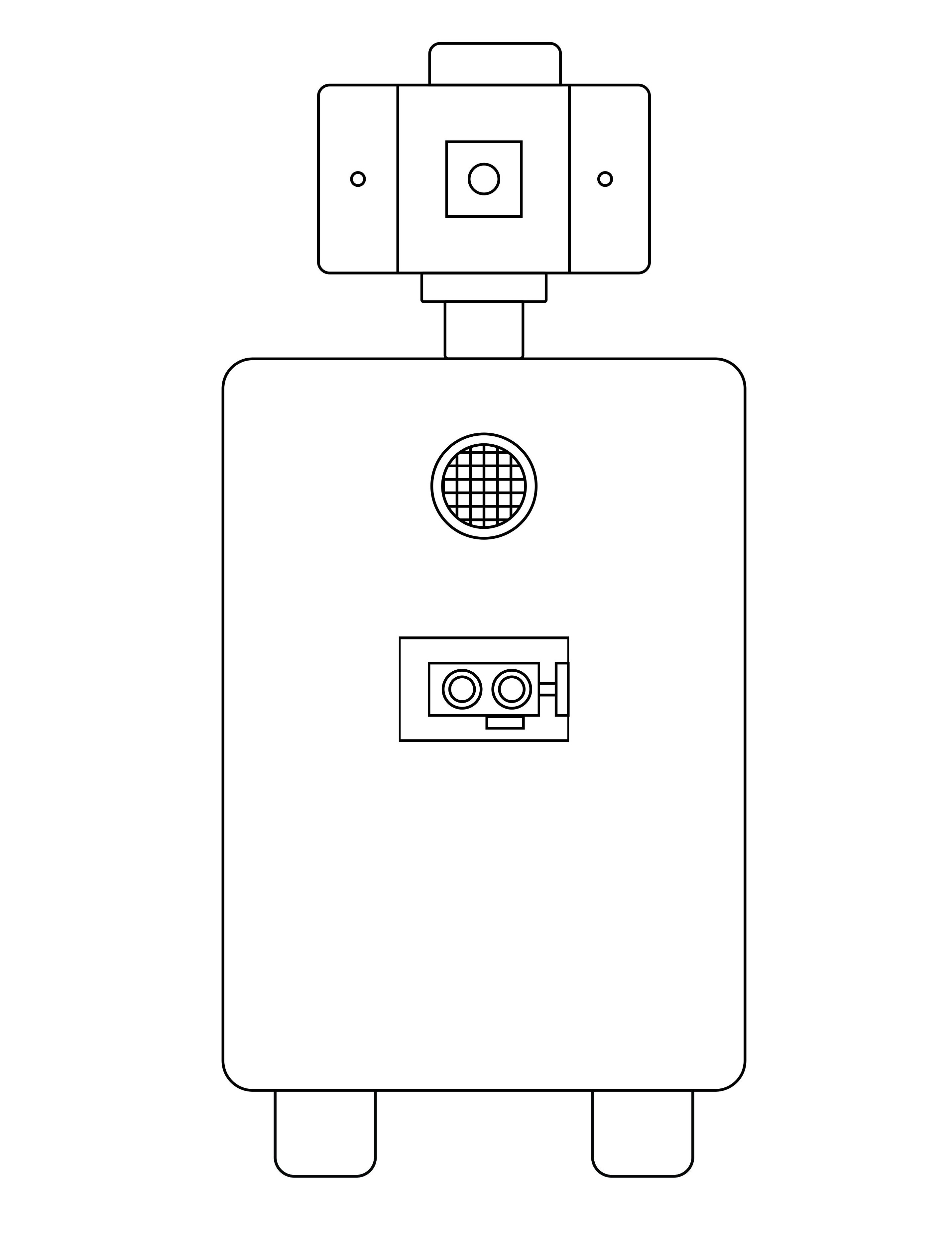}
	\caption{L-R. Left Sectional and Front View of Marvin}
	\label{MarvinFront}
\end{figure*}

\subsection{Locomotion Unit}
We utilize a trivial 4-wheel locomotion unit, this not only simplifies our initial prototype but also enables us to reduce its cost. This is controlled by the \emph{STM32 Nucleo F302R8T6} microcontroller running at 72kHz.\\
The microcontroller receives a 16-bit data packet over UART at a baud rate of 115200. The received data contains 10 bits of information consisting of the direction bits and the speed bits, this is received via interrupts and executed immediately by a self preempting high priority daemon. \newline

\begin{figure}[H]
	\centering
	\includegraphics[width=0.35\textheight]{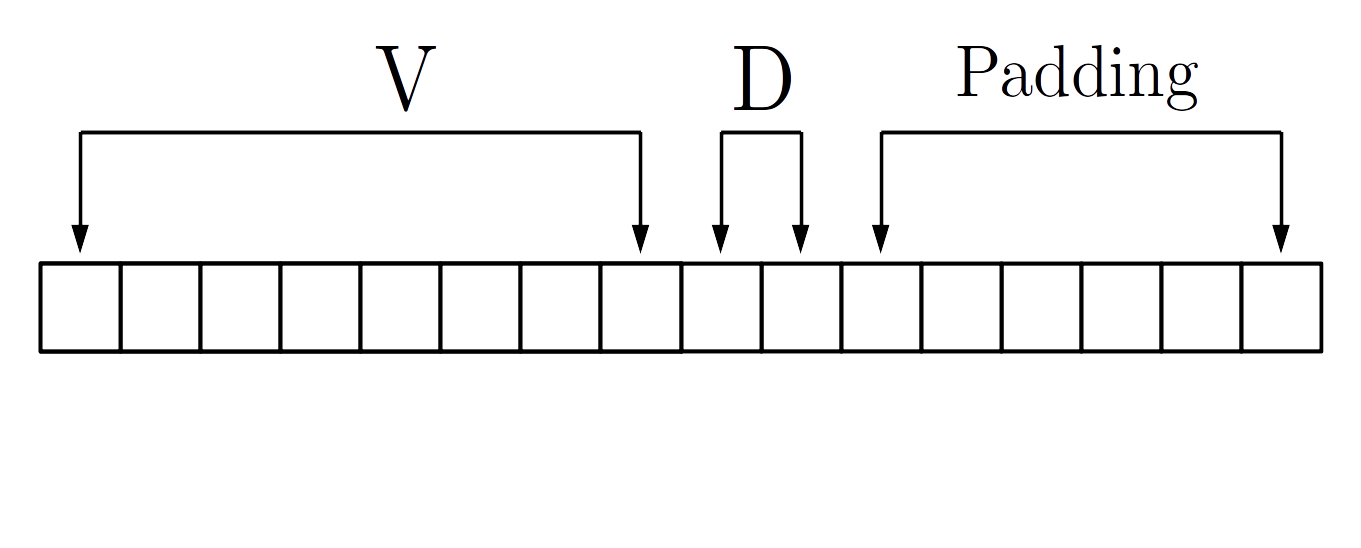}
	\caption{Locomotion Unit Control Data Packets}
	\label{LUC_Packet}
\end{figure}

\begin{description}
	\item[$\mathit{D}:$] 2 bits, represents the following directions:
	\begin{itemize}
		\item Left Forward
		\item Left Backward
		\item Right Forward
		\item Right Backward
	\end{itemize}
	\item[$\mathit{V}:$] 8 bits, represents speed (normalized to $\mathit{[0 , 255]}$).
\end{description}

The locomotion unit is stopped by setting the speed bits to zero in any direction.
The microcontroller then transmits the received data to the L293D motor controller via PWM pulses which in turn varies the voltages passed to the brushed DC motors.

\subsection{Obstacle Detection Unit} \label{ObstDetUnit}
We utilize a time-of-flight (ToF) based sensor for obstacle detection, namely the cost-effective HC-SR04 ultrasonic sensor which utilizes ultrasonic sound waves. We employ a simple beamforming technique, by rotating the sensor vertically with an MG996R brushed DC servo motor.
The maximum detectable obstacle distance is restricted to a threshold of 2.5$\mathit{m}$ to allow for faster sampling which is a common trade-off in ToF technologies that utilize slow flight mediums. The ideal obstacle distance can simply be computed from the time-of-flight as:

\begin{equation}
\mathit{D_{ideal} = \frac{C_{air} \times T}{2} }
\end{equation}

where $\mathit{D_{ideal}, C_{air}, T \in}  \mathbb{R} $, $\mathit{D_{ideal}}$ denotes flight distance (ignoring acoustic attenuation \cite{wiki:AcoAtt} of the sound wave,  $\mathit{C_{air}}$ denotes the speed of sound in air ($ \mathit{C_{air}} \approx 346 ms^{-1} $  at room temperature) and $\mathit{T}$ denotes time taken for the sound wave to reach an obstacle and return to the sensor. Thus, by limiting $\mathit{D_{ideal}}$ to a maximum of 2.5$\mathit{m}$, we can ideally obtain a sampling frequency of 69.2$\mathit{Hz}$.

The MG996R servo motor has a minimum operating latency of 0.14$\mathit{s}$ for every 60$^{o}$ at a running voltage of 6$\mathit{V}$ \cite{mg996r}, this can turn at a maximum angle of 120$^{o}$, therefore $\theta \in [0.0, 120.0]$. Though it does not have a feedback control system, future versions of the prototype will address this. There are more accurate servo motors with a larger $\theta$, but due to cost reasons, we utilized the MG996R servo motor

\begin{figure}[ht]
	\centering
	\setlength\fboxsep{0pt}
	\setlength\fboxrule{0.25pt}
	\includegraphics[width=0.3\textheight]{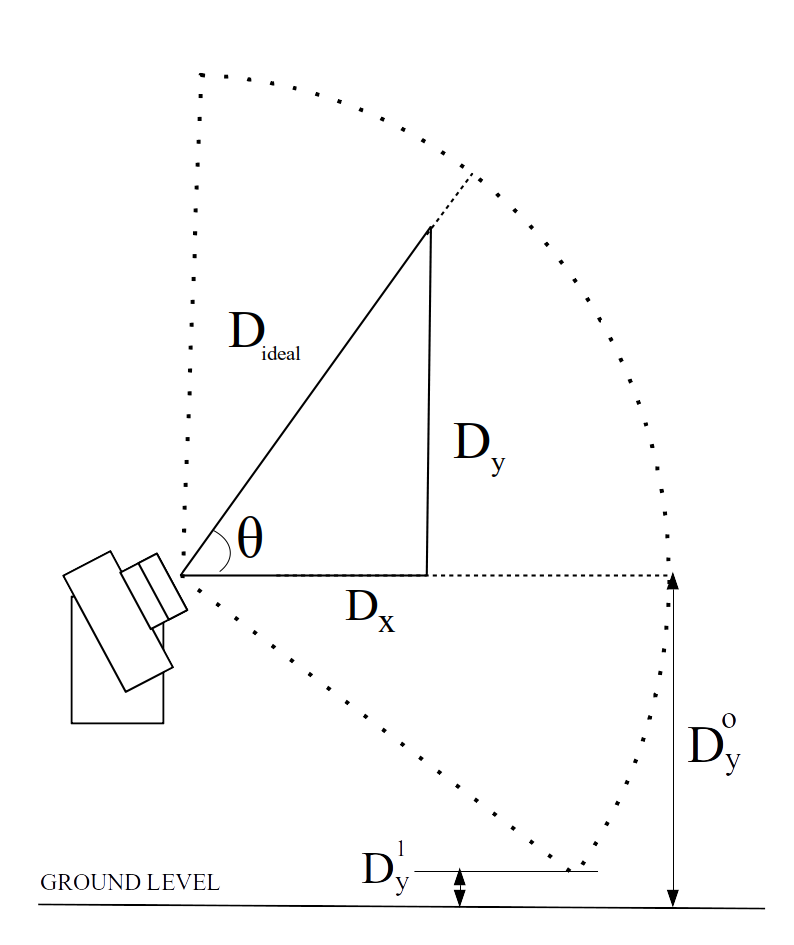}
	\caption{Obstacle Detection Technique using A Servo and an Ultrasonic Sensor}
	\label{BeamToF}
\end{figure}

Figure \ref{BeamToF} Illustrates how we perform beamforming by varying the inclination angle of the ultrasonic sensor $\mathit{\theta}$ through the servo motor. The unit is placed such that $\mathit{D_{y}^{1}}$ is small enough to be safely climbed by the robot and the 120$^{o}$ coverage of the ultrasonic servo covers our area of interest (the dotted arc).

Thus, the component distances relative to the ultrasonic can be computed using the simple relation:

\begin{equation}
\mathit{D_{y} = D_{ideal} \times \tan{\theta}}
\end{equation}

\begin{equation} 
\mathit{D_{x} = D_{ideal} \times \cos{\theta}}
\end{equation} 

While it is clear that the mechanism is highly dependent on the robot's body configuration, our evaluation also revealed that the mechanism is highly subject to environmental distortions, which can be compensated for using filtering algorithms. Most of our design decisions are greatly influenced by the limited hardware at our disposal, future versions hope to address this.

\subsection{Trunk}
The trunk has a cuboidal structure (See Figure \ref{MarvinFront}). This houses the batter/power pack, control circuitry, locomotion unit, and the STM32 Nucleo F302R8T6 microcontroller. Attached to its front is a mono-channeled USB speaker and the obstacle detection unit described in Subsection \ref{ObstDetUnit}.

\subsection{Head}
The \texttt{Head} houses the Raspberry Pi 3B microcomputer. Attached to its top is a microphone array used for sampling audio waveforms. To its front are attached two LEDs that illustrate the consciousness state of the robot i.e. Red for awaiting a command, Green for receiving a command, Yellow for executing a command. An RGB-888 0.5MP camera is also attached to its front for performing object detection, facial recognition, and other computer-vision-related tasks.
The \texttt{Head} is connected mechanically to the \texttt{Trunk} via the high-torque MG996R servo which tilts the \texttt{Head} and is controlled by the STM32 Nucleo F302R8T6 microcontroller.

\subsection{Cooling}
We perform forced air convection cooling by using heat sinks and high-velocity brushless 12$\mathit{V}$ and 5$\mathit{V}$ DC fans as the temperature of the system tend to rises to 80$^{o} C$ during prolonged usage.

\section {Keyword Spotting}\label{section:kws}
In this section, we discuss the Keyword Spotting system illustrated in Figure \ref{KwsArch}.
The Keyword Spotting System controls the \texttt{Latch}'s input to the \texttt{Semantic Interpreter} as not all sampled data needs to be processed by the \texttt{Semantic Interpreter}. 

This part of the system runs continuously on-device and detects for a Hotword. The major implication arising from this is that the Keyword Spotting (KWS) system needs to have minimal latency, enough to meet real-time requirements whilst maintaining decent false-accept and recall rates \cite{DBLP:journals/corr/SunRTPFMMSV17, SESRM}.

The Raspberry pi 3 typically runs at 1.2GHz (Quad-core) with a Broadcom BCM2837 64bit CPU having 1GB of RAM \cite{RPi}, this limits the intensity of computation and memory size of the model that can be run on this device, and thus the architectures need to be carefully evaluated and benchmarked.

Though Traditional systems employ Hidden Markov Models in KWS tasks to model both the keyword and background noise, they were superseded by Recurrent Neural Networks and Convolutional Neural Networks \cite{wiki:HMM, SESRM}. We utilize the \emph{cnn-trad-fpool-3} architecture which is one of the popular small-footprint keyword spotting architectures that have been increasingly gaining traction due to the limited number of multiplies and parameters they possess \cite{CnnKWS, DBLP:journals/corr/SunRTPFMMSV17, RnnT}.

\subsection{Dataset}
For our experiment, the word "Marvin" is chosen as the Hotword. We utilize the \emph{Environmental Noise Classification (ESC)} dataset \cite{esc50} to augment the training data and make the model rely on more robust features and be less susceptible to background noises.

We also used the \emph{Speech Commands Dataset} \cite{SpeechCommands}. Next, we manually crowd-sourced 569 audio samples from 125 volunteers within our university.
These datasets primarily consisted of single-channel waveforms at 48kHz sampled under quiet conditions. To reduce the computational load, we resampled the audio samples from 48kHz to 16kHz, at this sampling frequency, the intelligibility of human speech is retained and the discretized-time sampled signal also retains spatio-temporal information useful enough for inference \cite{wiki:Nyquist}.
Next, the resulting corpus is converted into a signed 16-bit PCM encoded format, clipped to 1s, then shuffled and split into training, test and evaluation sets in the ratio of 10:1:2 respectively.

\subsection{Neural Network Architecture}
The maximum duration $\mathit{D_{max}}$, allowed for an utterance of the keyword is $\mathit{1000 ms}$. Our implementation of \emph{cnn-trad-fpool-3} has an approximate size of 8.5 MB with an average running latency of 11.8 ms and a  $336 \mu s$ standard deviation per unit-batch inference, which is sufficient for our use case.

\begin{center}
	\begin{tabular}{ |c|c|c|c|c|c|c|c| } 
		\hline
		type & m & r & n & p & q & Par.  \\ 
		\hline
		conv & 24 & 10 & 64 & 1 & 3 & 15.4K  \\ 
		\hline
		conv & 12 & 5 & 64 & 1 & 1 & 164.8K \\ 
		\hline
		lin & - & - & 32 & - & - & 65.5K\\ 
		\hline
		dnn & - & - & 128 & - & - & 4.1K \\ 
		\hline
		softmax & - & - & 4 & - & - & 0.5K \\ 
		\hline 	\hline
		Total & - & - & - & - & - & 250.3K  \\ 
		\hline
	\end{tabular}
\end{center}
where:\\
$\mathit{m}$ = kernel size along the temporal dimension. \\
$\mathit{r}$ = kernel size along the frequency dimension. \\
$\mathit{n}$ = number of kernels or hidden units. \\
$\mathit{p}$ = downsampling stride size along the temporal  dimension. \\
$\mathit{q}$ = downsampling stride size along the spatial dimension.\\
This model architecture only pools in frequency and not along the temporal dimension.

\subsection{Augmentation}
To make the model more useful in real-world scenarios and more robust to noises, distortions, and compensate for informalities not reflecting in the original datasets, we applied a few augmentation techniques on the dataset, namely:
\begin{itemize}
	\item Pitch Augmentation
	\item Speed Augmentation
	\item Pitch-Speed Augmentation
	\item White-Noise Augmentation
	\item Background-Noise Augmentation
	\item Value Augmentation
	\item HPSS Augmentation
	\item Random Shift Augmentation
\end{itemize}
The noise augmentations were applied at SNRs randomly sampled between $\mathit{[-5dB, 10dB]}$

\subsection{Preprocessing}
To eliminate temporal redundancy in the sampled waveforms, we apply a 40-dimensional log-Mel energy filter bank feature extraction, this also makes training easier as many repetitive temporal features are removed in the process.
\subsection{Training}
\begin{figure*}[ht]
	\centering
	\setlength\fboxsep{0pt}
	\setlength\fboxrule{0.25pt}
	\includegraphics[width=0.3\textheight]{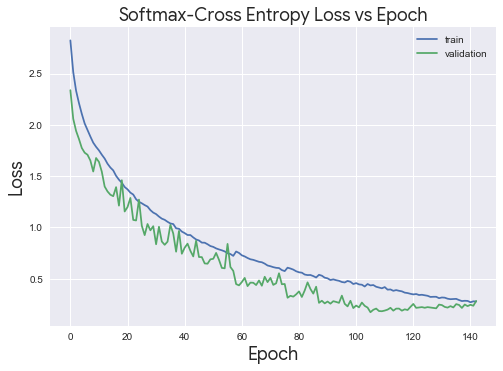}
	\includegraphics[width=0.3\textheight]{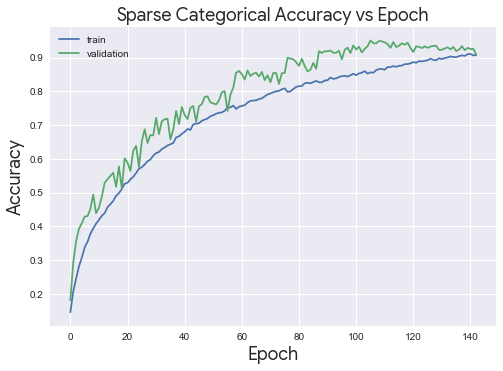}
	\caption{L-R Training Loss and Accuracy Graphs for Our \textbf{cnn-trad-fpool3} Implementation}
	\label{Losses}
\end{figure*}

A multi-category classification approach is employed to allow the model rely on more robust features of the input distribution, rather than in a binary classification scheme where the model is more liable to rely on non-robust features of the input distribution. The vanilla categorical cross-entropy loss function is used in this effect. \\
The learning rate and batch sizes are fine-tuned on the test, training, and evaluation sets.
Figure \ref{Losses} exhibits the training and test inferences' accuracy and loss graphs.

\begin{figure}[ht]
	\centering
	\setlength\fboxsep{0pt}
	\setlength\fboxrule{0.25pt}
	\includegraphics[width=0.2\textheight]{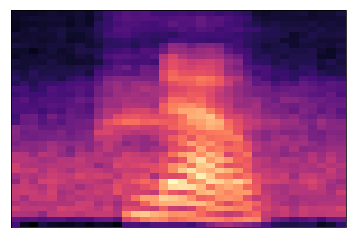}
	\caption{Sample Keyword Log-Mel Energy Filter Bank Feature}
	\label{figLMEFB}
\end{figure}

\subsection{Deployment}
To further reduce the computation load, we quantized the model from 32-bit floating-point precision to 8-bit signed-integer precision using the quantization regime specified in Subsection \ref{quantregime}, this comes with minimal accuracy tradeoffs.
We continuously shift the discretized-time sound sampling aggregation window by 250ms and aggregate 1s of samples, this enables us to have decent coverage of the incoming data samples as the position of the Hotword is non-deterministic. The aggregated sound samples are then passed to the interpreter to perform inference.

\section{Speech Synthesis} \label{section:speech}
The part of the system is used for generating natural-sounding speech and complements the \texttt{Control Skill} (illustrated in Figure \ref{KwsArch}) which responds to the user.

We utilized the Google Cloud Platform to synthesize natural-sounding high-fidelity speech via a pretrained generative autoregressive neural network, namely Wavenet \cite{wavenet, wiki:GCP}.
Wavenet especially takes advantage of dilated and casual convolutions which makes it efficient and model spatio-temporal features of speech at a remarkable accuracy.
This model generates natural-sounding audio whilst modeling linguistic features of a given text to a remarkable extent.
A downside to the model is its large size and inference latency which inherently made us deploy it to the Google Cloud Platform. With sufficient hardware upgrade, we can overcome this limitation.

\section{Intent Recognition} \label{section:intent}
This part of the system recognizes \texttt{Skill Intents} and \texttt{Tokens} from a given speech input and thus complements the \texttt{Semantic Interpreter} component of the architecture (See Subsection \ref{sub:sem}, Figure \ref{KwsArch}).
The intent recognition is performed using the Google Cloud platform's Dialogflow Speech-Intent engine \cite{wiki:Dflow} primarily because models yielding decent performance on intent recognition and/or natural language understanding tasks are often computationally intensive and have large memory requirements as the probabilistic distribution of the modeled data is highly multi-dimensional.
We utilized pre-existing DialogFlow \texttt{Skill} definitions and implemented custom ones such as Phone Call Initiation, Object Finding, and People Finding.

\section{Facial Recognition} \label{section:facial}
This part of the system is used for facial recognition, emotion detection, and intruder detection.
As such the system also needs to execute at a decent speed and accuracy, to this end, we utilize the populous Facenet neural network architecture \cite{FaceNet}.
Facenet performs at a very high accuracy on real faces ($99.65\%$ on the LFW dataset \cite{LFWTech}, $95.12\%$ on YouTube Faces Database \cite{YTFaces}. \\
Facenet's approach yields state-of-the-art performance on face-recognition tasks which is due to the great representational efficiency of the architecture whilst using only 128-bytes embeddings per face \cite{FaceNet}, with this approach we were able to eliminate the need to retrain our models every time a new face is added into the system as this would be very computationally intensive and impractical for real-world use cases. However, the saved 128-bytes embeddings are not used to train any additional classification model instead the face recognition module employs the use of the $L_{2}$ metric to calculate the distances between two face embeddings and identifies the subject.

The faces are recognized if the $L_{2}$ distance satisfies a specified unary predicate. The relation can be expressed as:

\[ S = \sum_{n=128}^{i=1} (\hat{E}_{i} - E_{i})^{2} \]
\[ IdentifierPredicate(S) = \begin{cases}
1,& \text{if } S\geq threshold\\
0,              & \text{otherwise}
\end{cases}
\]

where $\hat{E}_{i}, E_{i} \in \mathbb{R}^{128}, S \in \mathbb{R}^{1}$ and \\
$\hat{E}_{i}$ = Face Embedding\\
$E_{i}$ = Reference Face Embedding\\
$S$ = Squared Euclidean Distance Between $\hat{E}_{i}$ and $E_{i}$\\

We made use of the serially layered-pipeline approach and at the first layer of the pipeline the image taken from the camera is passed through a Haar Feature-Based Cascade Classifier \cite{wiki:haar_feature}. The Haar Feature-Based Cascade Classifier gives a bounding box to each of the faces detected in the image, the closest face to the camera (or the face with the biggest bounding box) is then sent down through the pipeline for further processing. At the second layer, the output of the first layer is resized to $\mathbb{R}^{1\times160\times160\times3}$, which is the tensor input shape of the Facenet model. At the third layer, the output of the second layer is normalized to a mean of 128 and standard deviation of 128 and then converted from 32-bit floating precision to it's unsigned 8-bit equivalent which is the expected input type of the quantized Tflite model, the output of this layer is then passed to the Tflite Facenet model which then computes the feature encoding ($\in \mathbb{R}^{128}$). We employed the use of quantization-aware training in other to minimize accuracy degradation when the model is converted to a quantized format.
To improve the model's speed and also reduce its size, we quantized it using the symmetric 8-bit signed integer quantization with the zero-point set to 0. The 8-bit signed quantization scheme approximates floating-point values using the relation \cite{TfQuant}:

\begin{equation} \label{quantregime}
real\_value = (int8\_value - zero\_point) \times scale 
\end{equation}

\section{Object Detection} \label{section:object}
This part of the system is intended for detecting objects and people, it is used for the Object Finding, Visually Impaired Guidance and other vision-related tasks described in Section \ref{sec:capab}. This is capable of detecting common items such as Kettles, Cups, Chairs, Beds, Tables, Laptops, Books, etc. and enables the robot to have visual cues which are especially useful for the visually impaired. We utilize a pretrained Mobilenet model \cite{howard2017mobilenets}.

The Mobilenet architecture is a streamlined and small-footprint architecture that performs remarkably well in mobile and embedded vision applications. The model is lightweight and highly efficient due to its extensive utilization of depthwise-separable convolutions which are a form of factorized vanilla convolutions \cite{howard2017mobilenets} whilst still modeling the data distribution remarkably though with a minimal accuracy tradeoff. This thus greatly reduces the number of multiplies and the model's size relative to a vanilla convolutional neural network. 
The model was quantized using the quantization regime described in \ref{quantregime}.

\section{Hardware Specifications} \label{section:hardware}

\begin{itemize}
	\item Acrylic Plastic: Used for the robot's physical enclosure. This was elected due to inaccessibility to 3d Printers.
	\item  Raspberry Pi 3B: 1.2GHz, Quad-core Broadcom BCM2837 64bit CPU, 1GB RAM \cite{RPi}. This is the main computation powerhouse. This has a minimal computational power that is good enough for the initial prototype.
	\item  Raspberry Pi Camera: 0.5MP, RGB-888. For capturing images used in object detection and facial recognition.
	\item STM32 Nucleo-F302R8T6: 72MHz, 16KB SRAM, 32 KB Flash \cite{STNucleo}. For controlling the obstacle detection unit and other physical components.
	\item HC-SR04 Ultrasonic Sensor: For obstacle detection.
	\item MG996R Brushed DC Servo Motor: $120^{o}$ range, 0.14s/60$^{o}$ @ 6V \cite{mg996r}. This tilts the ultrasonic sensor during beamforming.
	\item L293D Dual H-bridge DC Motor Driver.
	\item Generic Mono-channelled 16-bit USB Microphone: For keyword spotting and speech interaction.
	\item Generic USB Speaker: For audio and speech output.
	\item 74HC-595 8-bit Shift Register: For parallel-series output.
	\item 5V and 12V Brushless DC Fans: For forced-air convection cooling of the whole system.
	\item Heat Sinks: For convection cooling of the driver chips, microcontrollers, and microcomputers.
\end{itemize}

\section{Conclusion and Future Work}
In the literature, we proposed a novel architecture called IPRA. We also illustrated how we can implement a variety of user interfaces using the multi-modal user interface and hardware agnostic architecture. We investigated some of the important design decisions leading to a cost-effective and interactive modular robot. 
As a next step to help the adoption, exploration, and advancement of this Architecture, we plan on releasing the framework and robotic assistant system implementation.
In future high-fidelity prototypes, we plan on implementing and integrating more useful features, integrating high-quality sensors and actuators, and upgrading our hardware to enable it to run the whole software fastly and locally without dependence on cloud services.
Further, we plan to extend our experiments to different environmental conditions in the wild and integrate filtering and sensor fusion algorithms.

\section{Acknowledgments}
We would like to thank Prof. R. G. Jimoh for his feedbacks on each draft of the paper. We would also like to thank Yusuf Oladejo from the University of Ilorin for his valuable inputs and feedback on this project.

%\includepdf[pages={1,2}]{references.pdf}
\bibliography{main.bib}{}
\bibliographystyle{plain}

\end{document}